%% file: sn-article.tex
\documentclass[sn-mathphys-num, iicol]{sn-jnl}

\usepackage{graphicx}%
\usepackage{multirow}%
\usepackage{amsmath,amssymb,amsfonts}%
\usepackage{amsthm}%
\usepackage{mathrsfs}%
\usepackage[title]{appendix}%
\usepackage{xcolor}%
\usepackage{textcomp}%
\usepackage{manyfoot}%
\usepackage{booktabs}%
\usepackage{algorithm}%
\usepackage{algorithmicx}%
\usepackage{algpseudocode}%
\usepackage{listings}%
\usepackage[utf8]{inputenc}
\usepackage{subcaption}

\theoremstyle{thmstyleone}%
\theoremstyle{thmstyletwo}%

\theoremstyle{thmstylethree}%

\begin{document}

\title[Article Title]{Self-Organizing Complex Networks with AI-Driven Adaptive Nodes for Optimized Connectivity and Energy Efficiency}


\author[1]{\fnm{Azra} \sur{Seyyedi}}\email{azra.seyyedi@iasbs.ac.ir}

\author[2]{\fnm{Mahdi} \sur{Bohlouli}}\email{me@bohlouli.com}

\author[1, 3]{\fnm{SeyedEhsan} \sur{Nedaaee Oskoee}}\email{nedaaee@iasbs.ac.ir}

\affil*[1]{\orgdiv{Department of Physics}, \orgname{Institute for Advanced Studies in Basic Sciences (IASBS)}, \orgaddress{\street{Prof Yousef Sobouti Blvd}, \city{Zanjan}, \postcode{45137-66731}, \state{Zanjan}, \country{Iran}}}

\affil[2]{\orgdiv{Research and Innovation Department}, \orgname{Petanux GmbH}, \orgaddress{\street{Dottendorfer}, \city{Bonn}, \postcode{53129}, \state{North Rhine-Westphalia}, \country{Germany}}}

\affil[3]{\orgdiv{Research Center for Basic Sciences and Modern Technologies (RBST)}, \orgaddress{\street{Prof Yousef Sobouti Blvd}, \city{Zanjan}, \postcode{45137-66731}, \state{Zanjan}, \country{Iran}}}


\abstract{High connectivity and robustness are critical requirements in distributed networks, as they ensure resilience, efficient communication, and adaptability in dynamic environments. Additionally, optimizing energy consumption is also paramount for ensuring sustainability of networks composed of energy-constrained devices and prolonging their operational lifespan.	
In this study, we introduce an Artificial Intelligence (AI)-enhanced self-organizing network model, where each adaptive node autonomously adjusts its transmission power to optimize network connectivity and redundancy while lowering energy consumption. Building on our previous Hamiltonian-based methodology, which is designed to lead networks toward globally optimized states of complete connectivity and minimal energy usage, this research integrates a Multi-Layer Perceptron (MLP)-based decision-making model at each node. By leveraging a dataset from the Hamiltonian approach, each node independently learns and adapts its transmission power in response to local conditions, resulting in emergent global behaviors marked by high connectivity and resilience against structural disruptions. This distributed, MLP-driven adaptability allows nodes to make context-aware power adjustments autonomously, enabling the network to maintain its optimized state over time. 
Simulation results show that the proposed AI-driven adaptive nodes collectively achieve stable complete connectivity, significant robustness, and optimized energy usage under various conditions, including static and mobile network scenarios. This work contributes to the growing field of self-organizing networks by illustrating the potential of AI to enhance complex network design, supporting the development of scalable, resilient, and energy-efficient distributed systems across diverse applications.	
}




\maketitle

\section{Introduction}\label{sec:introduction} 
\input{sections/introduction}

\section{Related Work} 
\label{sec:relared_work}
\input{sections/related-work}

\section{Methodology}
\label{sec:methodology}
\input{sections/methodology}

\section{Analysis of Results and Discussion}
\label{sec:results}
\input{sections/results}

\section{Conclusion and Future Works}
\label{sec:conclusion}
\input{sections/conclusion}
\bibliography{sections/bibliography-paper3}

\end{document}

%% file: sections/introduction.tex
In distributed networks, achieving high connectivity and robustness is essential for enabling resilience, efficient communication, and adaptability in dynamic environments. 
Robustness ensures high connectivity despite failures, while resilience reflects the network's capacity to adapt and recover from disruptions \cite{zhang2015notion} \cite{gao2015recent}.
Distributed networks, where each node operates independently and without centralized control, play a crucial role in fields such as sensor networks \cite{qi2001distributed}, emergency response \cite{lopez2018distributed}, Internet of Things (IoT) ecosystems \cite{alsboui2021distributed}, and autonomous systems \cite{wu2024distributed}. At the same time, energy consumption remains a critical concern, particularly as many of these networks consist of energy-constrained devices with limited battery life \cite{cardei2005improving}. Thus, optimizing energy usage without compromising network performance is crucial to ensure the longevity and sustainability of these systems. 

Traditional methods for ensuring connectivity and robustness often rely on static configurations or centralized control, which may not be suitable for large-scale, distributed, or mobile networks. In contrast, self-organizing networks offer a promising approach by allowing nodes to autonomously form connections and maintain an optimized network state through local interactions, particularly when combined with advanced decision-making algorithms. Recent advancements in AI have opened new possibilities for enhancing the adaptability of network nodes, allowing them to autonomously optimize their behavior to improve overall network performance \cite{sayed2014adaptive}. This study focuses on advancing self-organizing complex networks through the integration of AI, specifically in the form of adaptive nodes capable of independently learning optimal behaviors.

Existing research on self-organizing networks has explored various methods for achieving a globally optimized network structure, often drawing on mathematical principles or heuristic approaches. In our prior work \cite{seyyedi2023energy}, we proposed a Hamiltonian-based methodology that enabled nodes to autonomously adjust their transmission power and form links based on transmission range and spatial proximity with neighbors. This approach guided the network toward a globally stable state with near-complete connectivity, high robustness, and minimized energy consumption. Through this Hamiltonian framework, the network demonstrated a capacity for self-sustaining, optimized topology even against local or significant disruptions. While being effective, this approach relied on a predefined, numerical method, leaving potential for further adaptability and resilience through learning-based techniques.

In this study, we introduce an AI enhanced methodology that leverages deep learning to enhance node adaptability within the self-organizing complex network. Specifically, each node is equipped with two MLP models trained on a dataset generated from our Hamiltonian-based approach. This MLP models allow each node to autonomously learn optimal transmission power adjustments and also link creation based on local conditions, making context-aware decisions that collectively drive the network toward global objectives of connectivity, robustness, and energy efficiency. By decentralizing decision-making to each node, we create a fully distributed, adaptive system where network optimization emerges from the coordinated, AI-driven actions of individual nodes. This approach not only extends the network’s self-organizing capability but also allows it to dynamically respond to environmental changes, maintaining optimal performance over time.

This paper, at first, introduces a distributed method for optimizing connectivity and energy efficiency in distributed networks by embedding a decision-making process at the node level. Then, presents a data-driven adaptation mechanism for autonomous transmission power adjustments and link status decision, demonstrating that MLP-based learning enhances network resilience and adaptability. Finally, provides simulation results that illustrate how AI-driven adaptive nodes, functioning in a self-organizing complex network, achieve superior connectivity, robustness, resilience, and energy efficiency in various scenarios compared to conventional approaches. 
This study demonstrates the effectiveness of this approach in both static and mobile network environments. 
It highlights the potential of integrating deep learning within complex network design to build scalable, resilient, and energy-efficient distributed systems for a wide range of real-world applications.

The remainder of this paper is organized as follows. Section \ref{sec:relared_work} provides an overview of existing related work. Section \ref{sec:methodology} outlines the methodology, presenting the details of the proposed approach, including the adaptive AI-driven network model and its implementation. Section \ref{sec:results} focuses on the analysis of results and discussion, showcasing various plots and figures under different scenarios, such as static and mobile networks, two-dimensional and three-dimensional topologies, and scenarios with and without node failures. Finally, Section \ref{sec:conclusion} concludes the paper by summarizing the findings and highlighting potential directions for future research.

%% file: sections/related-work.tex
The study of self-organizing complex networks has attracted significant attention in recent years, particularly in the context of distributed systems, wireless sensor networks (WSNs), and the IoT. These networks are characterized by the autonomous formation and adaptation of connections among nodes based on local interactions, enabling robust and efficient communication in dynamic environments.

Early foundational work in self-organizing networks focused on the principles of decentralized control and local interaction. For instance, Barabási and Albert \cite{barabasi1999emergence} introduced the concept of scale-free networks, which exhibit self-organization through preferential attachment, allowing for resilient network structures that maintain connectivity despite node failures. Their work laid the groundwork for understanding how complex networks can evolve over time based on local rules.

Authors in \cite{ghanea2006self} further explored the dynamics of a self-organizing network, emphasizing the importance of distributed local decision-making in establishing global network resilience. This concept has been applied in various contexts, including mobile ad-hoc networks (MANETs), where nodes dynamically form links based on proximity and transmission power \cite{banerjee2020self}. 
Recent studies including \cite{diaz2019review} have investigated self-organizing behaviors in WSNs as well, where the optimization of communication protocols has become critical for enhancing network lifetime and efficiency. \textit{Ye et al} in \cite{ye2016survey} provided a comprehensive review of self-organization methods in distributed networks of agents, underscoring the significance of self-organizing mechanisms in achieving sustainable operation.

Adaptive control mechanisms are crucial for enabling nodes to respond to changing network conditions. Research \cite{ismat2019adaptive} proposed an adaptive protocol for mobile WSNs, focusing on the dynamic adjustment of transmission power based on local topology and energy constraints.
Advancements in machine learning (ML) and AI have further enhanced the adaptability of nodes within self-organizing networks. \textit{Zhao et al} \cite{zhao2021deep} introduced a reinforcement learning approach for optimizing resource allocation in WSNs, allowing network to manage both power and time for throughput maximization. The simulation results demonstrate that the proposed transmission policies can produce higher throughput in the local network and finally improve overall system performance in comparison with greedy policy, random policy and conservative policy.

In our previous work \cite{seyyedi2023energy}, we demonstrated a Hamiltonian-based methodology for optimizing connectivity and robustness in self-organizing networks, where nodes autonomously adjust their transmission power to achieve a stable state with near-complete connectivity and low energy usage. This foundational work informs the current study by providing a framework upon which AI-driven adaptive mechanisms can be integrated.

Despite the progress in self-organizing networks and AI-driven methodologies, there remains a gap in fully realizing the potential of adaptive nodes in achieving real-time, distributed decision-making that optimally balances connectivity and energy efficiency. This study addresses this gap by introducing an AI-based approach that allows nodes to autonomously learn optimal power adjustments based on local network conditions, enhancing the adaptability and resilience of self-organizing complex networks. Through the integration of ML and Hamiltonian optimization, this work aims to establish a robust framework for future research and applications in distributed network design.

%% file: sections/methodology.tex
The methodology employed in this study is centered on the development and evaluation of a self-organizing complex network model with AI-driven adaptive nodes. The primary objectives are to achieve maximum (100\%) connectivity, ensure significant robustness and resilience, and optimize energy usage under varying network conditions. This section elaborates on the dataset generation process, the adaptive learning framework, and the simulation setup used to analyze network performance in both static and mobile scenarios across 2D and 3D environments.

\begin{figure}[!h] 
	\centering
	\includegraphics[width=0.5 \textwidth]{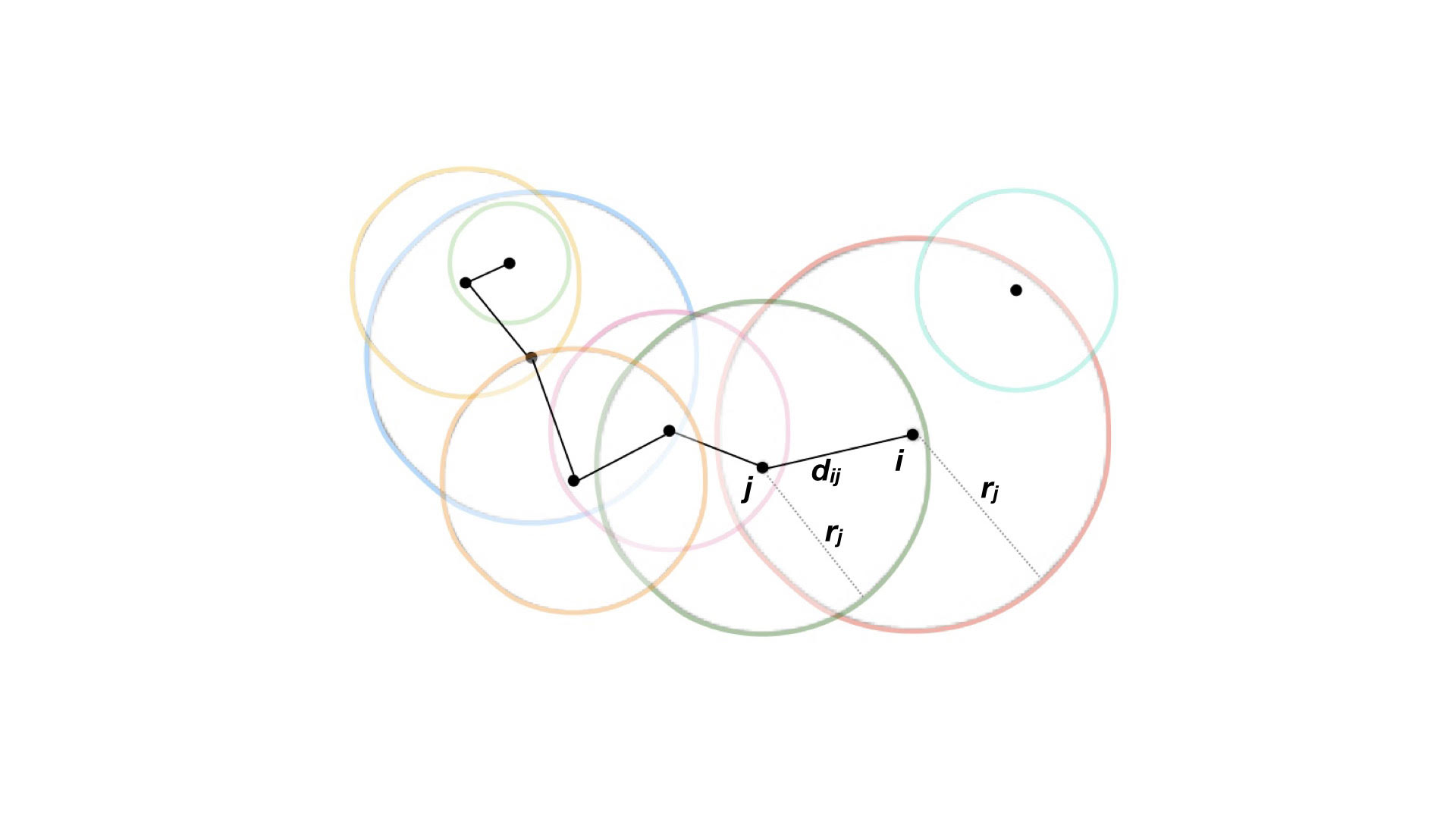}
	\caption{Illustration of adaptive network nodes and their connections based on Euclidean distance and transmission range, focusing on a specific spatial configuration within the network. Nodes are homogeneous in capabilities, and links are established between any two nodes if they are within each other’s transmission range \cite{seyyedi2023energy}.}
	\label{fig:concept_image}
\end{figure}

Our network model consists of a distributed complex network of  100 adaptive nodes, each capable of adjusting transmission power to dynamically form or dissolve links with neighboring nodes. Nodes are homogeneous in capabilities, with each node equipped with a limited transmission range that can be modified. Links between nodes are established when they are within each other’s transmission range, with connection governed by Euclidean distance (Figure \ref{fig:concept_image}). The network operates in a fully distributed, ad-hoc manner, with nodes functioning autonomously and without centralized control or predetermined routing protocols. 
The physical and channel characteristics of nodes in this study are considered within standard operating ranges and are not the focus of analysis. The primary emphasis is placed on the node's ability to adapt transmission power and connectivity autonomously to optimize network performance.

Nodes can move within a bounded square area, leading to dynamic network topology.
Each node is aware only of its local neighborhood (i.e., nodes within its transmission range). Power adjustments impact both the transmission range and energy consumption of nodes, with a trade-off between transmission range and energy efficiency. Network stability is defined by a global state of maximum connectivity, robustness to disruptions, and optimized energy consumption.

To provide each node with a dataset for AI-driven decision-making, we first employed a Hamiltonian-based optimization framework from our previous work to model an optimized network state. In this approach, the Hamiltonian function 
($H$) is designed to minimize energy usage while maximizing connectivity and robustness across the network. Each node decides adjustments to its transmission power based on this function, with the goal of steering the network toward a stable state that meets the connectivity, robustness, and energy efficiency conditions.

The network should maintain near-complete connectivity, with each node reachable from any other through a series of links. The network should sustain connectivity even in the presence of local failures or node removals. Nodes should operate at minimal energy expenditure, achieved through careful power control.
Using this Hamiltonian approach, we generated a dataset comprising optimal transmission power values for various network states. This dataset includes local network topology information, energy consumption metrics, and power adjustments required to achieve optimized network conditions. The Hamiltonian formulation is designed to minimize energy consumption while maximizing connectivity, providing an optimal baseline for training the AI model.

Building on the dataset generated from the Hamiltonian framework, we designed an MLP model to enable each node to make real-time, adaptive transmission power adjustments, enabling nodes to independently predict the optimal transmission power based on local network conditions. Another MLP model is trained to make decisions about link establishment when the conditions for it exist.
To decide whether it is better to accept the link if another node comes within its range and has connection conditions. So, These MLP models serve as the decision-making core of each adaptive node. Each MLP model is trained on a subset of the dataset relevant to local network conditions, allowing nodes to learn the relationship between network topology, power settings, and desired outcomes.
 
The first MLP architecture consists of an input layer, three hidden layers (optimized based on validation results), and an output layer. 
The input features to this MLP include degree of the node which is a connectivity metric of neighboring nodes and node's current transmission range.
The output of this MLP determines the transmission power adjustment for each node, ensuring a balance between connectivity and energy efficiency. 
The second MLP architecture is the same. It consists of an input layer, three hidden layers, and an output layer. For each node, second MLP model takes as input a vector of local parameters, including the node’s current transmission range, distance to desired node, local connectivity (node degree), T, which is metric assigned to each node representing its freedom to affect the status of its communications. This model predicts the updated state of the link between itself and the node under consideration.

The dataset generated by the Hamiltonian-based framework is used to train both MLP models, with training and validation datasets split to prevent overfitting.
Each node operates independently, using its trained MLP models to continuously adapt its transmission power and manage connections. 
%
Through repeated application of this process, this distributed AI-driven approach allows nodes to self-organize and adapt dynamically to varying environmental and network conditions, demonstrating emergent global behaviors such as high connectivity and resilience against disruptions.

To assess the effectiveness of the AI-driven self-organizing network, we evaluate network performance based on the following metrics:

\begin{itemize}
	\item \textbf{Connectivity Ratio:} The percentage of nodes connected within a single giant component of the network, reflecting the network’s overall connectedness.
	\item \textbf{Robustness:} Measured by the network’s resilience to node failures, defined as the ability to maintain connectivity when a percentage of nodes or links are removed.
	\item \textbf{Energy Consumption:} The average energy usage per node, which captures the network’s efficiency in power management. The energy consumed of node devices considered inversely related to the square of the distance as $ power \propto \frac{1}{r^2}$, based on the \textit{Friis transmission equation} \cite{friis1946note}.
	\item \textbf{Link Count Ratio:} A metric used to evaluate the connectivity of a network by comparing the actual numbers of links in the network to the total possible number of links. It provides a measure of how densely the nodes in a network are connected. 
	\item \textbf{Adaptability:} The network’s ability to reorganize and maintain optimal performance when faced with local disturbances, such as node mobility or varying density.
\end{itemize}

The proposed model was evaluated through extensive simulations conducted in Python. The simulations were designed to compare network performance across static and mobile nodes in both 2D and 3D network environments. Key parameters included:

\begin{itemize}
	\item \textbf{Node Distribution densities:} Simulations were conducted for varying densities to assess the scalability of the model.
	\item \textbf{Mobility dynamics:} Random movement patterns were introduced for mobile nodes, with speeds sampled from a normal distribution.
	\item \textbf{Node failures:} Scenarios were simulated with up to 50\% node failures to test the resilience of the network.
\end{itemize}
Each simulation tracks critical performance metrics. For mobile networks, the model is expected to demonstrate adaptability by responding to dynamic topology changes, with nodes recalibrating their transmission power based on real-time local conditions. The outcomes confirm that the proposed method achieves stable and maximized connectivity while maintaining low energy usage.

%% file: sections/results.tex
\begin{figure*}[htbp]
	\centering
	\begin{subfigure}{\textwidth}
		\centering
		\includegraphics[width=\textwidth]{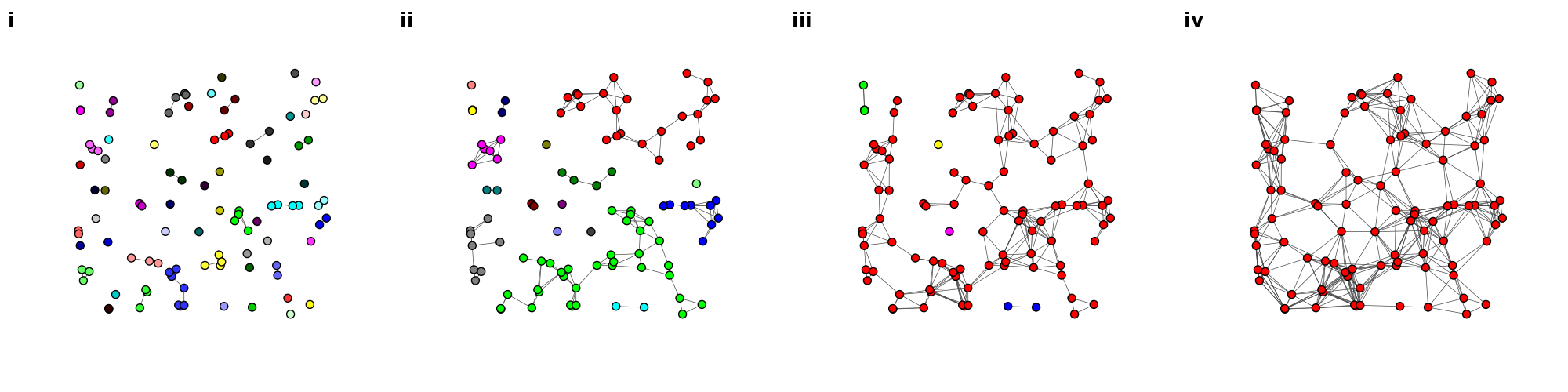}	
		\caption{Static network snapshots at steps 0, 200, 500, and 2500, showing the merging of components over time.}
		\label{fig:local-combined-snapshots}
	\end{subfigure}
	\vspace{0.5cm}
	\begin{subfigure}{\textwidth}
		\centering
		\includegraphics[width=\textwidth]{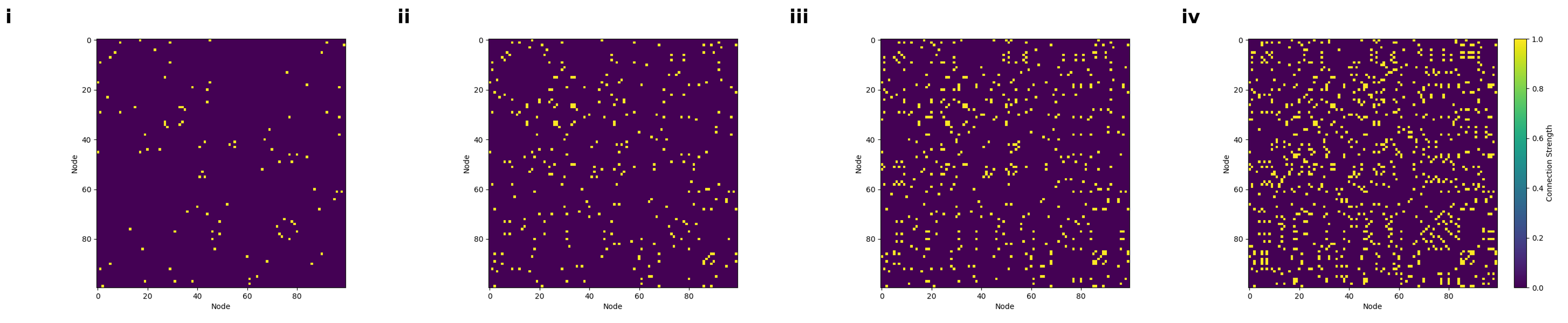}	
		\caption{Connectivity matrices for static network snapshots, illustrating connections between nodes at each stage.}
		\label{fig:local-connectivity-matrices}
	\end{subfigure}
	\vspace{0.5cm}
	\begin{subfigure}{\textwidth}
		\centering
		\includegraphics[width=\textwidth]{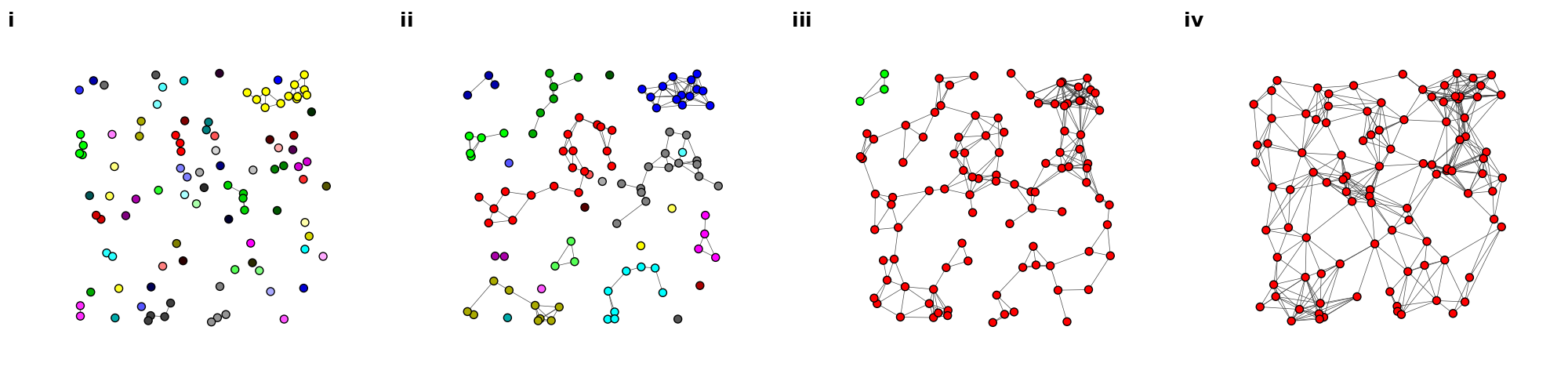}	
		\caption{Mobile network snapshots at steps 0, 200, 500, and 2500, showing component merging with node mobility.}
		\label{fig:movement-combined-snapshots}
	\end{subfigure}
	\vspace{0.5cm}
	\begin{subfigure}{\textwidth}
		\centering
		\includegraphics[width=\textwidth]{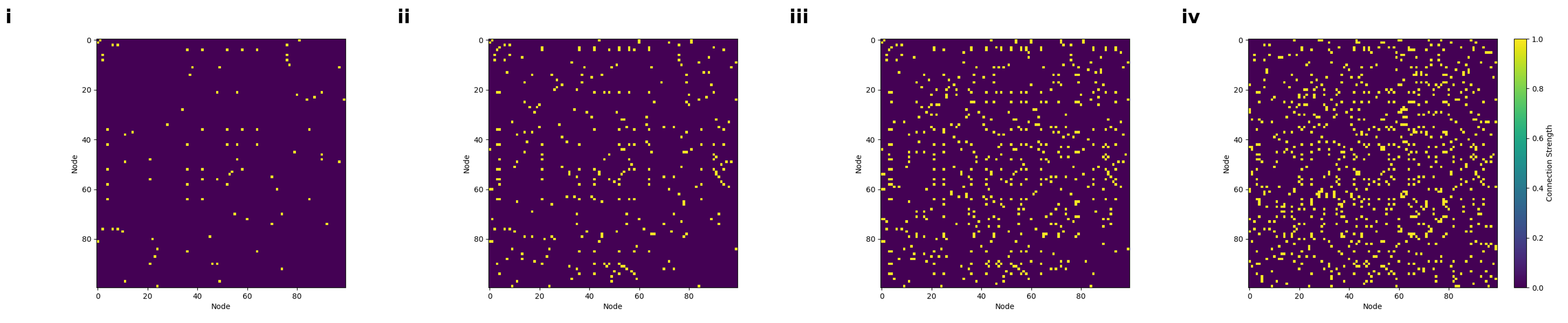}	
		\caption{Connectivity matrices for mobile network snapshots, showing evolving connections as nodes move.}
		\label{fig:movement-connectivity-matricess}
	\end{subfigure}
	\caption{Snapshots and connectivity matrices of the complex network of adaptive nodes at steps 0, 200, 500, and 2500. (a) Static network snapshots showing the evolution of node connectivity and the merging of components over time until reaching full connectivity, with nodes displayed in different colors to distinguish components. (b) Connectivity matrices for static network snapshots, where yellow cells indicate connections between node pairs. (c) Mobile network snapshots, similar to (a), showing node connectivity with movement introduced over time. (d) Connectivity matrices for mobile network snapshots, reflecting network progression with node movement. The networks are spatial, with x and y coordinates considered in the snapshots (a) and (c). Connectivity matrices (b) and (d) are symmetric, reflecting undirected, bidirectional links.}
	\label{fig:snapshots}
\end{figure*}

This section begins by analyzing the network's performance in terms of the evaluation metrics under varying conditions.
At first, some snapshots of dynamic behavior and connectivity progression in our adaptive networks with both static and mobile nodes over time are illustrated in \ref{fig:snapshots}. 
Subfigures \ref{fig:local-combined-snapshots} and \ref{fig:local-connectivity-matrices} display evolution of adaptive network with static nodes and subfigures \ref{fig:movement-combined-snapshots} and \ref{fig:movement-connectivity-matricess} offer an analogous progression for the mobile adaptive network. 
Subfigures \ref{fig:local-combined-snapshots} and \ref{fig:movement-combined-snapshots} show snapshots of the networks at different steps related to various stages. In each snapshot, nodes are represented based on their spatial coordination. They display node connectivity at different steps, with colors distinguishing separate components to highlight merging behavior as the network evolves. Each snapshot shows that, initially, multiple components are present, which gradually merging into a single connected component, achieving full connectivity. By step 2500, all nodes belong to a single component (indicated by uniform color), confirming full connectivity with fewer links than a fully saturated network, which highlights an efficient, adaptive structure.

Subfigures \ref{fig:local-connectivity-matrices} and \ref{fig:movement-connectivity-matricess} show the corresponding connectivity matrices for each step, with yellow cells representing connected node pairs. These matrices provide an additional view into the network’s structural evolution, where increasing yellow areas reflect the growing interconnectivity of nodes. The initial sparsity of yellow cells at step 0 indicates limited connections, while the dense yellow region at step 2500 reflects a cohesive, fully connected network. Symmetry in the matrices reflecting the undirected, bidirectional nature of the connections.
In \ref{fig:movement-combined-snapshots}, It seems that the mobility of nodes facilitates faster component merging. The connectivity matrices in \ref{fig:movement-connectivity-matricess} reinforce this observation, sounds that node mobility contributes to a more rapid filling of connections over time, demonstrated by an earlier emergence of dense yellow regions compared to the static network.

These visualizations underscore the adaptive and efficient connectivity mechanisms in both static and mobile networks. This efficiency is key in adaptive networks, as it demonstrates that both types of networks achieve maximum connectivity with low link redundancy. This characteristic is beneficial for reducing overall network energy usage, while ensuring that all nodes remain connected under dynamic conditions.

\begin{figure*}
	\centering
	\begin{subfigure}[b]{0.48\textwidth}
		\includegraphics[width=\textwidth]{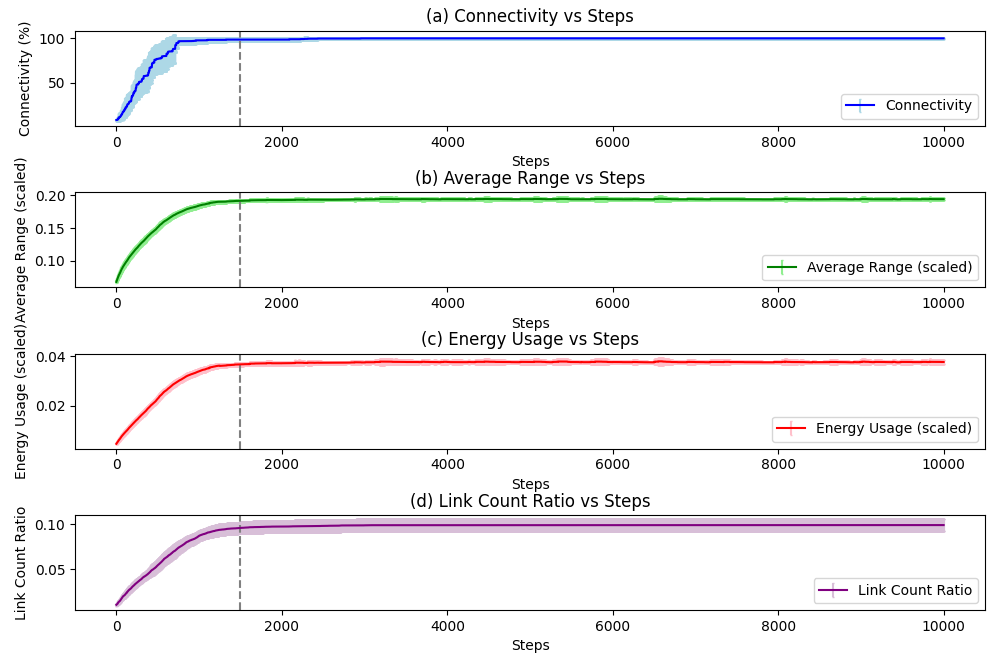}
		\caption{(i) Demonstrates the transition from initial low connectivity to 100\% connectivity, achieved as the network self-organizes. (ii) Shows the evolution of scaled transmission range, which increases and stabilizes as the network achieves full connectivity. (iii) Indicates energy consumption based on the scaled range following the Friis transmission equation. Energy stabilizes as connectivity becomes stable. (iv) Displays the ratio of active links to total possible links, reaching a stable value as the network self-organizes.}
		\label{fig:local-connectivity-density=0.05}
	\end{subfigure}
	\hfill
	\begin{subfigure}[b]{0.48\textwidth}
		\includegraphics[width=\textwidth]{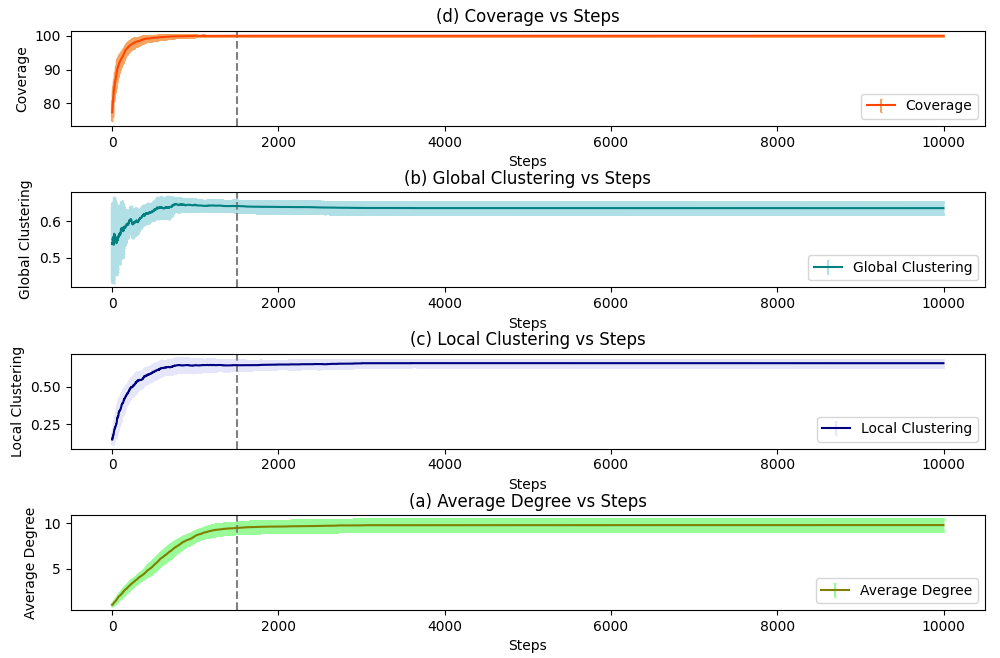}
		\caption{(i) Shows the extent of network coverage, which improves and stabilizes as the network reaches 100\% connectivity. (ii) Tracks global clustering, indicating network cohesion as it stabilizes at high connectivity. (iii) Tracks local clustering, showing neighborhood clustering as connectivity is established. (iv) Shows the network's average degree, increasing and stabilizing as connectivity is achieved. \vspace{2\baselineskip}}
		\label{fig:local-coverage-density=0.05}
	\end{subfigure}
	
	\vskip\baselineskip  
	
	\begin{subfigure}[b]{0.48\textwidth}
		\includegraphics[width=\textwidth]{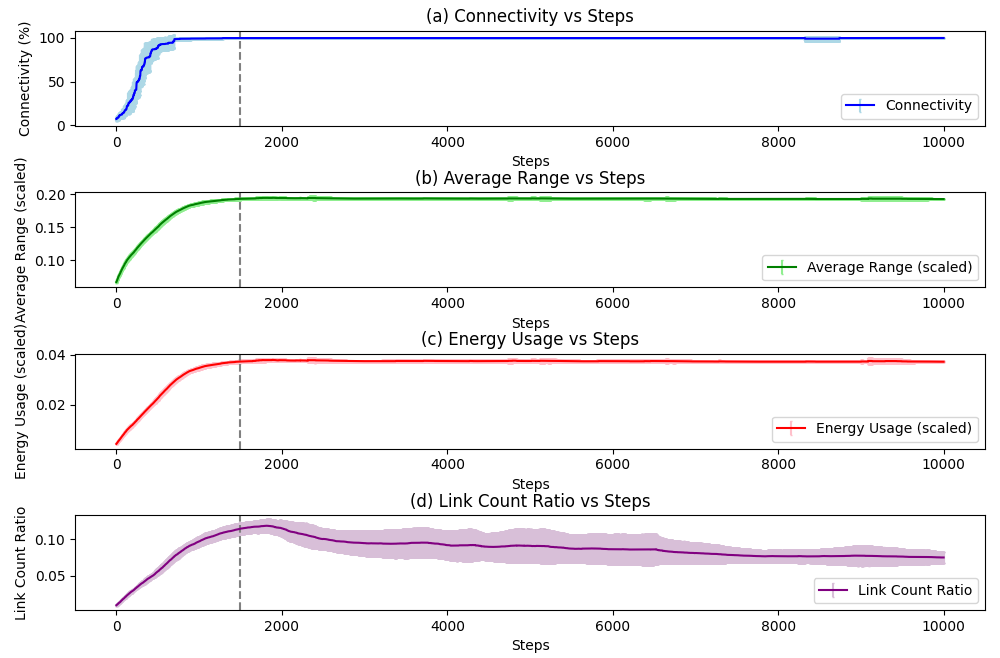}
		\caption{Similar to (a), but with nodes in motion. As mobility influences link dynamics, the network shows adaptive connectivity while maintaining efficient link usage and energy consumption.}
		\label{fig:movement-connectivity-density=0.05}
	\end{subfigure}
	\hfill
	\begin{subfigure}[b]{0.48\textwidth}
		\includegraphics[width=\textwidth]{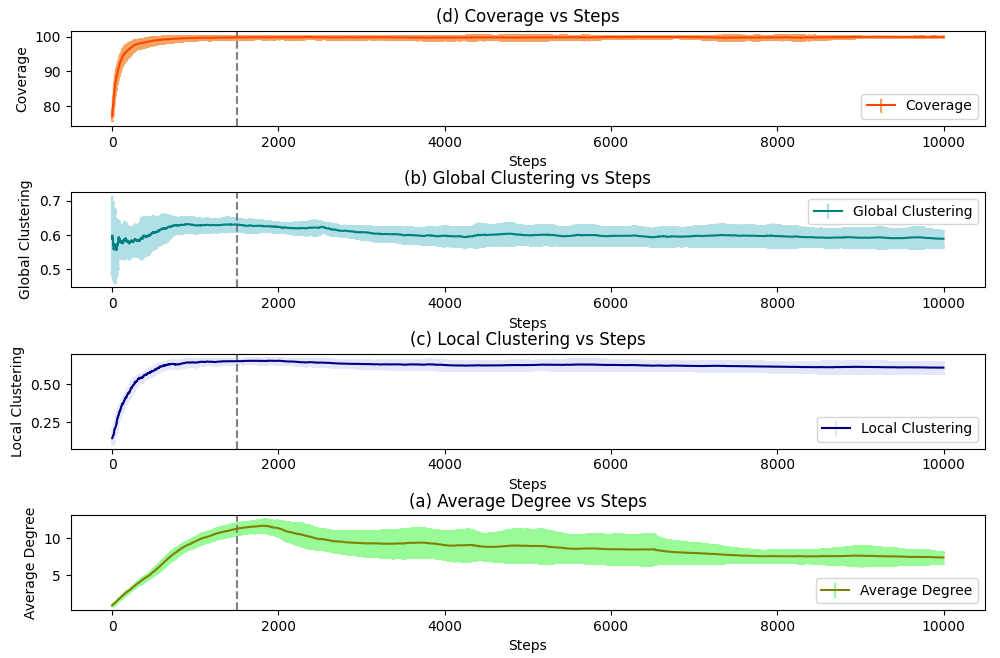}
		\caption{Similar to (b), but with mobile nodes, emphasizing how mobility affects clustering, coverage, and average degree in a dynamic network. \vspace{1\baselineskip}}
		\label{fig:movement-coverage-density=0.05}
	\end{subfigure}
	
	\caption{This figure presents the evolution of key network metrics over 10,000 steps, averaged over 10 different initial configurations for each network type. The error bars reflect the variability among these simulations, capturing the robustness of the results despite differences in initial node configurations. Subfigures (a) and (b) depict metrics for the complex network of static adaptive nodes, while (c) and (d) show equivalent metrics for mobile adaptive nodes, all at a node distribution density of 0.05. The metrics tracked include connectivity, average range, energy usage, link count ratio, coverage, global clustering, local clustering, and average degree. The data reveals how both static and mobile networks reach stable connectivity and other desired states, with mobile networks adapting under the challenge of random movement dynamics.}
	\label{fig:connectivity-and-convergence}
\end{figure*}

In the following, figure \ref{fig:connectivity-and-convergence}, we present an in-depth analysis of dynamics of the adaptive complex networks with both static and mobile nodes over a long simulation time (over 10,000 steps), providing insight into network behaviors at different stages. The evaluation is conducted using key performance metrics and supporting indicators, providing detailed insights into the network's behavior over time. Each metric is analyzed and and discussed comprehensively to highlight the network's performance. 
In each experiment, the network's behavior was averaged across 10 simulations with identical parameters but different initial configurations to ensure statistical reliability. This averaging approach ensures that the results are not biased by any single starting configuration, providing a more comprehensive picture of network behavior. The error bars in each plot indicate the variability among these initial configurations, showcasing the consistency and reliability of the connectivity patterns and other metrics achieved by the adaptive network algorithm.
 
Each metric highlights different aspects of the network's performance.
Notable points are included in the following.
Connectivity Progression:
In both static \ref{fig:local-connectivity-density=0.05} and mobile \ref{fig:movement-connectivity-density=0.05} networks, initial connectivity starts low and progressively, reaches 100\% around step 1500. Beyond this point, the connectivity stabilizes and maintains a consistent value, showing successful network adaptation to full connectivity in a short time and its ability to sustain this almost complete connectivity over time across different runs. This rapid increase suggests that nodes efficiently adjust their transmission ranges to form a single-component network, regardless of initial configuration. This consistency across diverse starting conditions highlights the adaptability of the network model in achieving connectivity.

For the mobile adaptive nodes, random movement dynamics introduce further variability; however, the network still reaches stable states, evidenced by the convergence patterns and error bar ranges. This suggests that the adaptive mechanism effectively compensates for node movement, maintaining a fully connected network. However, the slight fluctuations observed indicate periodic adjustments as nodes move in and out of communication range.

Average Range and Energy Usage:
The scaled transmission range in \ref{fig:local-connectivity-density=0.05} and \ref{fig:movement-connectivity-density=0.05} increases as the network seeks connectivity but stabilizes once 100\% connectivity is achieved. This stability minimizes further range adjustments, conserving energy.
Energy usage, derived from the transmission range, in \ref{fig:local-connectivity-density=0.05} and \ref{fig:movement-connectivity-density=0.05} stabilizes after reaching full connectivity, showing the balance between connectivity and energy efficiency.
The initial increase in energy usage aligns with the growth phase in connectivity, as nodes expand their range to establish links. As connectivity stabilizes, energy usage also plateaus, demonstrating that nodes no longer need to increase their range, leading to efficient energy usage across different runs.
Mobile nodes, challenged by random movement, still maintain low energy use while adapting their range to compensate for movement-induced topology changes.

Link Count Ratio:
The link count ratio \ref{fig:local-connectivity-density=0.05} indicates the fraction of active links relative to possible links. A stable ratio reflects efficient use of links, avoiding unnecessary connections even when full connectivity is achieved. It follows trends similar to connectivity and stabilizes at a low value. Once the network achieves full connectivity, further link formation is unnecessary. As the network reaches such state of equilibrium, new connections and disconnections balance dynamically to maintain stable connectivity. This indicates that the network achieves connectivity efficiently, with much fewer links than a fully connected graph, thus conserving energy and reducing complexity. 

The mobile network \ref{fig:movement-connectivity-density=0.05} maintains a similar pattern, demonstrating adaptability with minimal links and without excessive redundancy, regardless of starting conditions.
It stabilizes at a level comparable to the static network, though with visible fluctuations reflecting the transient connections formed as nodes move. This fluctuation indicates that while the network achieves a high degree of connectivity, link stability is periodically disrupted and re-established due to mobility.

Coverage: Coverage is crucial as it reflects the spatial reach of the network over the monitored area. Subfigures \ref{fig:local-coverage-density=0.05} and \ref{fig:movement-coverage-density=0.05} reveal that both networks eventually can fully cover the domain area once connectivity stabilizes, reflecting efficient spatial adaptation. For mobile nodes, the slight variability due to random movement dynamics reflects occasional gaps in coverage, which are quickly compensated as the network adapts.

Clustering Coefficients:
Clustering measures the tendency for nodes to form tightly knit groups. The high global clustering in both networks reflects a robust structure with redundancy, which enhances fault tolerance. Local clustering shows that neighboring nodes are also interconnected, adding stability at the local level. Global and local clustering coefficients in \ref{fig:local-coverage-density=0.05} and \ref{fig:movement-coverage-density=0.05} indicate network cohesion. These high clustering values demonstrate that the network balances connectivity and redundancy, making it more resilient to disruptions. Mobility slightly reduces clustering in \ref{fig:movement-coverage-density=0.05}, as moving nodes periodically alter local neighborhoods.

Average Degree:
The average degree demonstrated in \ref{fig:local-coverage-density=0.05} and \ref{fig:movement-coverage-density=0.05} rises with connectivity but stabilizes at a moderate value, reflecting the network's optimization in connecting nodes without excessive energy expenditure. Higher degree values would imply more energy usage, while lower values could reduce robustness. By achieving a balanced degree, the network maximizes connectivity while reducing unnecessary links.

In comparing static and mobile networks, both types reach similar stable states, but the mobile network exhibits slightly more variability, as indicated by broader fluctuated error bars. This reflects the challenges due to changing topology introduced by movement, where nodes constantly adjust their transmission ranges to preserve connectivity. This random movement approach  with varying velocities provides a realistic stress test for the network’s ability to self-organize and maintain connectivity autonomously.
The early rapid adjustments followed by stabilization across all metrics confirm that, even with different initial configurations, the network self-organizes to achieve full robust connectivity with minimal, efficient transmission ranges.
Overall, These plots collectively highlight the adaptive algorithm's effectiveness in achieving a balance between connectivity and energy efficiency, even when faced with random dynamics.

\begin{figure*}[h!]
	\centering
	\begin{subfigure}{\textwidth}
		\centering
		\includegraphics[width=\textwidth]{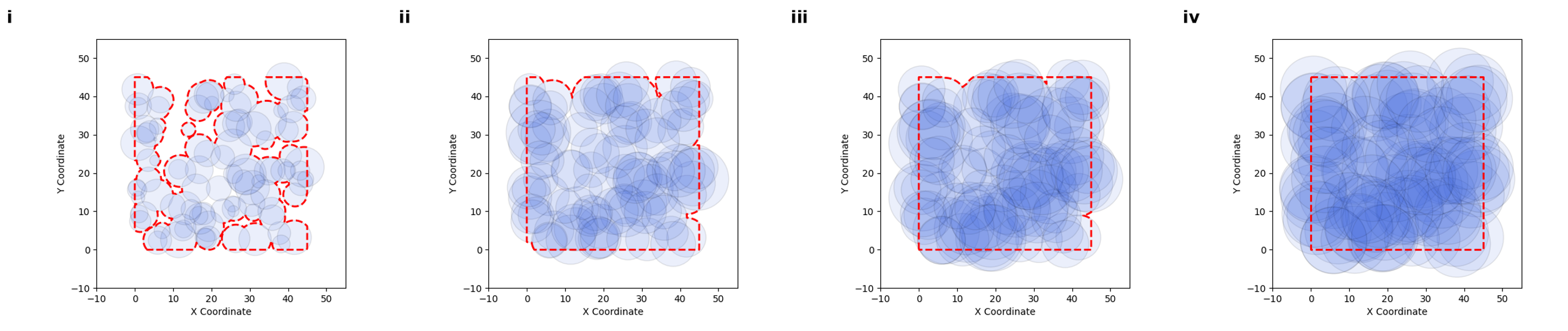}	
		\caption{Coverage snapshots for static adaptive nodes.}
		\label{fig:local-coverage-combined-snapshots}
	\end{subfigure}
	\vspace{0.5cm}
	\begin{subfigure}{\textwidth}
		\centering
		\includegraphics[width=\textwidth]{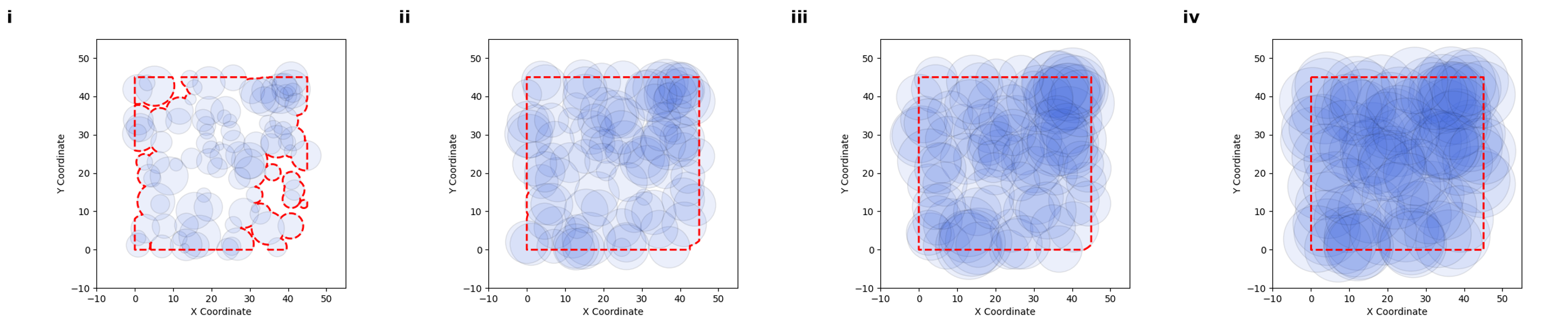}	
		\caption{Coverage snapshots for mobile adaptive nodes.}
		\label{fig:movements-coverage-combined-snapshots}
	\end{subfigure}
	\caption{This figure illustrates the evolution of network coverage over time for two types of adaptive networks: static (a) and mobile (b). The snapshots are taken at steps 0, 200, 500, and 2500, demonstrating how each network adapts its coverage area as nodes adjust their transmission ranges. The blue circles represent the transmission range of individual nodes, while the red dashed outline shows the boundary of the total covered area. In both cases, the networks begin with partial coverage and progressively expand their reach, achieving full coverage.}
	\label{fig:coverage}
\end{figure*}	

In this phase of the network’s process, we examine the spatial coverage of adaptive nodes at various stages. Figure \ref{fig:coverage} visualizes the spatial coverage snapshots of static and mobile networks over time. 
As shown in \ref{fig:local-coverage-combined-snapshots}, the nodes initially cover small, isolated areas within the network, as indicated by scattered blue transmission circles within the area. Over time, as nodes adaptively adjust their transmission ranges, the coverage regions expand and increasingly overlap, indicating enhanced connectivity and robustness in the network. The red dashed lines, representing the coverage boundary, reveal the spatial reach of the network over time. 

Mobile network \ref{fig:movements-coverage-combined-snapshots} compared to the static network, seems to achieve broader and more consistent coverage in fewer steps. This resilience makes the mobile network suitable for dynamic environments, where maintaining coverage in the face of movement is crucial.
Achieving stable boundaries demonstrates that both networks can sustain a high-quality spatial reach.
This observation ensures full network coverage besides supporting the overall goals of the study. The achievement of full coverage occurs autonomously, driven by the adaptive mechanisms embedded in the network's design.  
This snapshots illustrate that the adaptive algorithm enables both networks to achieve extensive coverage while balancing stability (in the static network) and adaptability (in the mobile network).

\begin{figure*}[h!]
	\centering
	\begin{subfigure}{\textwidth}
		\centering
		\includegraphics[width=\textwidth]{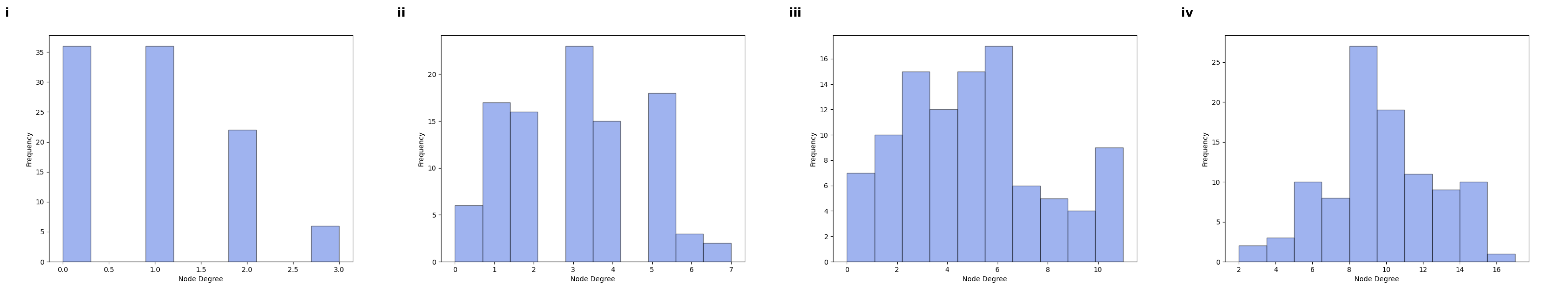}	
		\caption{Degree distribution histograms for the static adaptive network at time steps 0, 200, 500, and 2500. }
		\label{fig:local-degree-combined-histograms}
	\end{subfigure}
	\vspace{0.5cm}
	\begin{subfigure}{\textwidth}
		\centering
		\includegraphics[width=\textwidth]{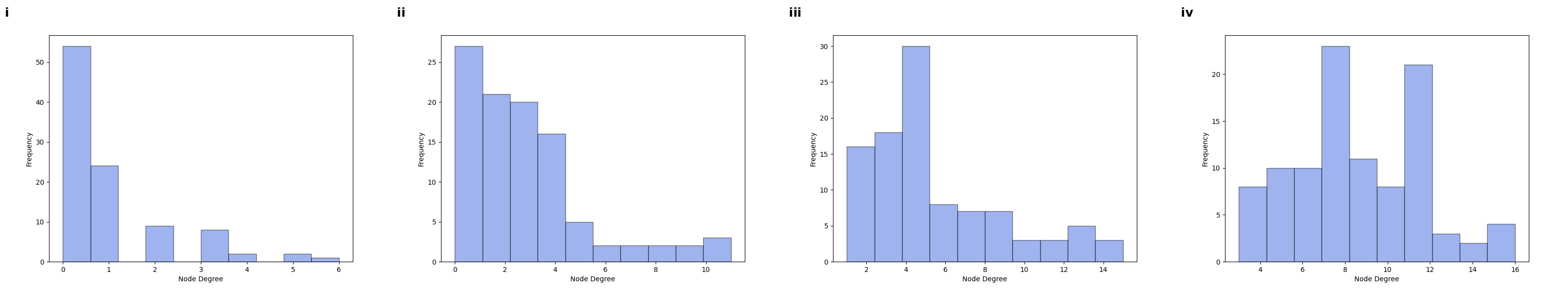}	
		\caption{Degree distribution histograms for the mobile adaptive network at time steps 0, 200, 500, and 2500.}
		\label{fig:movement-degree-combined-histograms}
	\end{subfigure}
	\caption{Degree distribution histograms of the adaptive complex network, comparing static and mobile node scenarios at different time steps (0, 200, 500, and 2500). The histograms show the evolution of node connectivity (node degree) as the adaptive algorithm increases network connectivity while maintaining energy efficiency. Subfigure (a) represents the static network, while subfigure (b) represents the mobile network, illustrating the impact of node mobility on the degree distribution over time.}
	\label{fig:degree-histogram}
\end{figure*}	

The histograms in figure \ref{fig:degree-histogram} illustrate the evolution of node degree distribution over time in a complex network of adaptive nodes. These degree distribution histograms provide insights into the adaptive process of the network. At the initial step (i) shown in \ref {fig:local-degree-combined-histograms} , the network begins with a relatively low and uneven degree distribution, reflecting a starting configuration where nodes have yet to optimize their connectivity. As the steps progress (ii-iii), the network undergoes self-organization through adaptive adjustments in each node’s transmission range, leading to changes in the degree distribution. By step 2500 (iv), the degree distribution stabilizes, achieving a broader range of node degrees with a higher average degree than the initial state. This analysis supports the concept of an adaptive mechanism to balance connectivity with potential energy costs. 

Mobile networks \ref{fig:movement-degree-combined-histograms} exhibit a similar trend but with greater variability due to movement dynamics. 
However, random movement introduces additional variability, seen in the degree distribution’s less consistent shape across snapshots. This variation indicates the dynamic challenges of maintaining connectivity under mobility.
The evolution of these degree distributions aligns well with the goals of maximizing connectivity and robustness. For both static and mobile scenarios, the adaptive algorithm effectively manages connectivity, achieving a adequate degree count without excessive redundancy, supporting efficient energy use. The results provide evidence that the algorithm efficiently handles both scenarios, validating the network's adaptability and robustness.

\begin{figure*}[h!]
	\centering
	\begin{subfigure}[b]{0.48\textwidth}
		\centering
		\includegraphics[width=\textwidth]{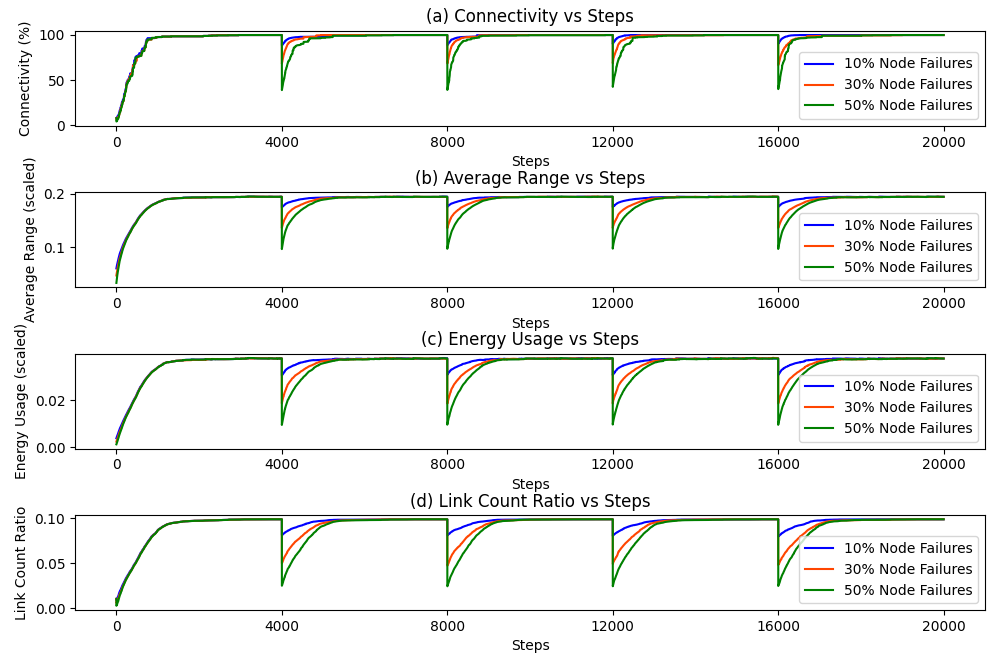}	
		\caption{Performance of Static Network under Node Failures. }
		\label{fig:local-failures}
	\end{subfigure}
	\hfill
	\begin{subfigure}[b]{0.48\textwidth}
		\centering
		\includegraphics[width=\textwidth]{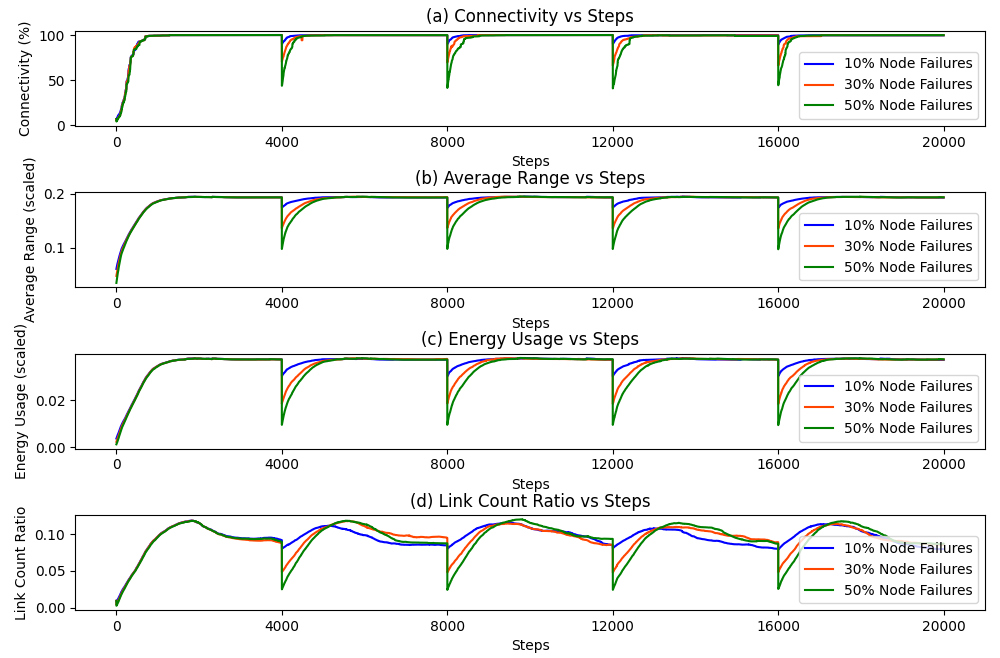}	
		\caption{Performance of Mobile Network under Node Failures.}
		\label{fig:movemnt-failures}
	\end{subfigure}
	\caption{Adaptive Network Performance under Node Failures in Static and Mobile Configurations. (a) Shows results for a static network with node failures, while (b) illustrates a mobile network. Each subfigure presents the evolution of connectivity, average transmission range, energy usage, and link count ratio over simulation steps under scenarios of 10\%, 30\%, and 50\% node failures (blue, orange, and green lines, respectively). The adaptive mechanism allows both networks to recover connectivity over time.}
	\label{fig:failures}
\end{figure*}	

To simulate realistic scenarios and assess the network's resilience, we incorporate node failures as part of the evaluation framework. Node failures serve to model situations in which some nodes become non-functional, either due to hardware issues, battery depletion, or intentional attacks, and thus temporarily lose their ability to communicate within the network. Unlike traditional simulations where failed nodes are permanently removed from the network, our approach includes a recovery mechanism that allows affected nodes to rejoin the network by recalculating and adjusting their transmission range based on the current network conditions. By allowing nodes to reintegrate, we aim better observe the network's capacity to recover from disruptions and maintain critical connectivity metrics. 

Figure \ref{fig:failures} examines network performance under failures of 10\%, 30\%, and 50\% of nodes. In both scenarios, displayed in \ref{fig:local-failures} and \ref{fig:movemnt-failures}, connectivity initially drops upon failure but recovers as nodes dynamically adjust their ranges. Energy usage temporarily increases during recovery but stabilizes efficiently. Two other metrics also show the similar trends.
The analysis demonstrates that the network exhibits a strong capacity for self-healing and restoration of connectivity when subjected to significant node failures (up to 50\% of nodes). This self-organizing behavior emphasizes the network’s robustness, as it consistently re-establishes full connectivity even under significant disruptions and adverse conditions. These findings support the utility of the proposed adaptive mechanism in applications requiring reliable and energy-efficient connectivity under dynamic conditions.

\begin{figure*}[h!]
	\centering
	\begin{subfigure}[b]{0.48\textwidth}
		\centering
		\includegraphics[width=\textwidth]{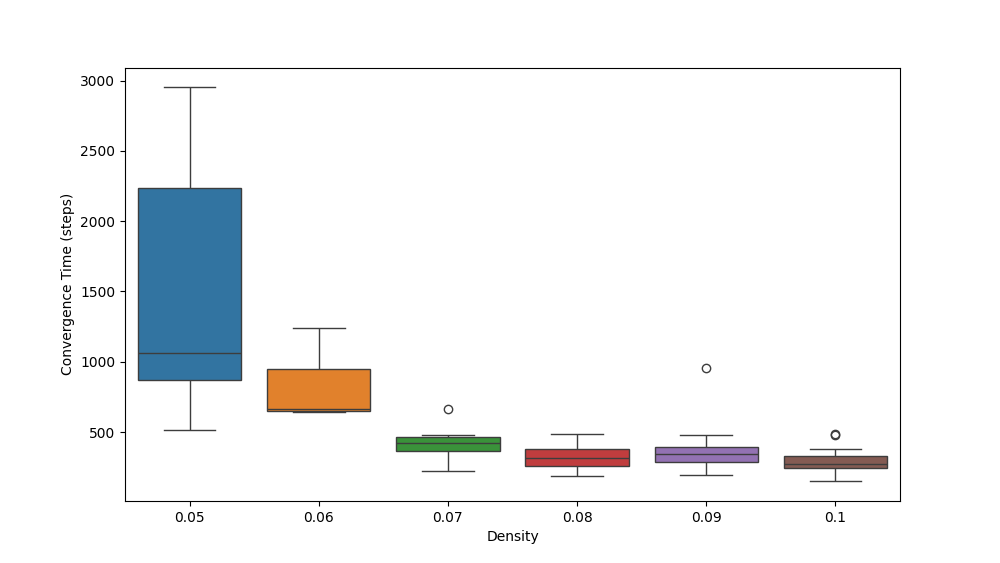}	
		\caption{Convergence time vs. node distribution density under static network conditions. The plot shows a decrease in convergence time as node density increases, with lower densities exhibiting greater variability in convergence times. }
		\label{fig:local-convergence-density}
	\end{subfigure}
	\hfill
	\begin{subfigure}[b]{0.48\textwidth}
		\centering
		\includegraphics[width=\textwidth]{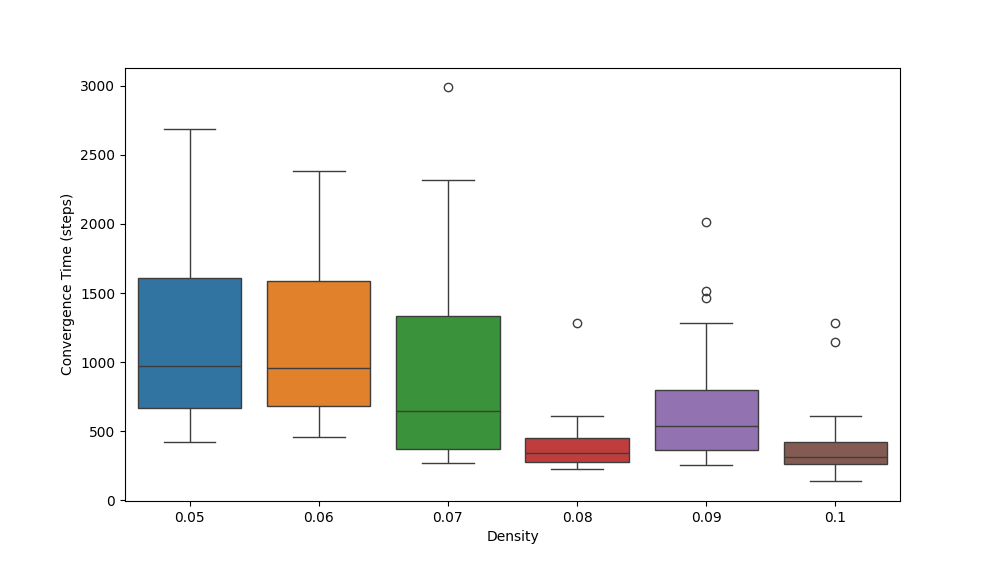}	
		\caption{Convergence time vs. node distribution density under mobile network conditions. Convergence times also decrease with increasing density; however, mobility introduces variability and outliers, indicating less predictable convergence behavior.}
		\label{fig:movement-convergence-density}
	\end{subfigure}
	\caption{Convergence time versus node distribution density for static and mobile network conditions. Each box plot shows the distribution of convergence times across different density levels, illustrating the impact of node density on network connectivity dynamics under local (a) and movement-induced (b) conditions.}
	\label{fig:convergence-density}
\end{figure*}	
 
All the figures and simulations presented so far correspond to networks with a distribution density of 0.05. However, the convergence time required to achieve full connectivity may be influenced by node distribution density in the area. To investigate this potential dependency, figure \ref{fig:convergence-density} is included to analyze and determine whether such a relationship exists.
In the static network, (See figure \ref{fig:local-convergence-density}) convergence time decreases remarkably with increasing density, demonstrating that a denser distribution of nodes enables quicker establishment of network-wide connectivity. Lower densities, particularly at 0.05 and 0.06, exhibit a larger spread in convergence times, reflecting greater variability in the network’s ability to converge. This may be due to isolated nodes or smaller clusters that prolong the time required to integrate all nodes into a single connected component. In contrast, at higher densities (0.08 and above), the convergence times become both shorter and more consistent, indicating that increased node density promotes stability and efficiency in connectivity formation.

In the mobile network, \ref{fig:movement-convergence-density}, a similar trend of decreasing convergence time with higher density is observed; however, the results show increased variability across all density levels. At moderate densities (e.g., 0.07), convergence times are widely spread, suggesting that node mobility introduces additional challenges to the convergence process. Movement dynamics can lead to temporary disconnections and reconnections, adding unpredictability to the time required for the network to reach full connectivity. This effect is evident in the larger spread of data points and the presence of outliers across various densities, highlighting that while mobility may offer flexibility, it complicates the process of establishing stable, connected network states.
Overall, these findings suggest that while increasing density generally improves convergence times in both static and mobile networks, mobility requires further consideration due to its potential to disrupt network stability, particularly at lower and moderate densities.

\begin{figure}[!h] 
	\centering
	\includegraphics[width=0.5 \textwidth]{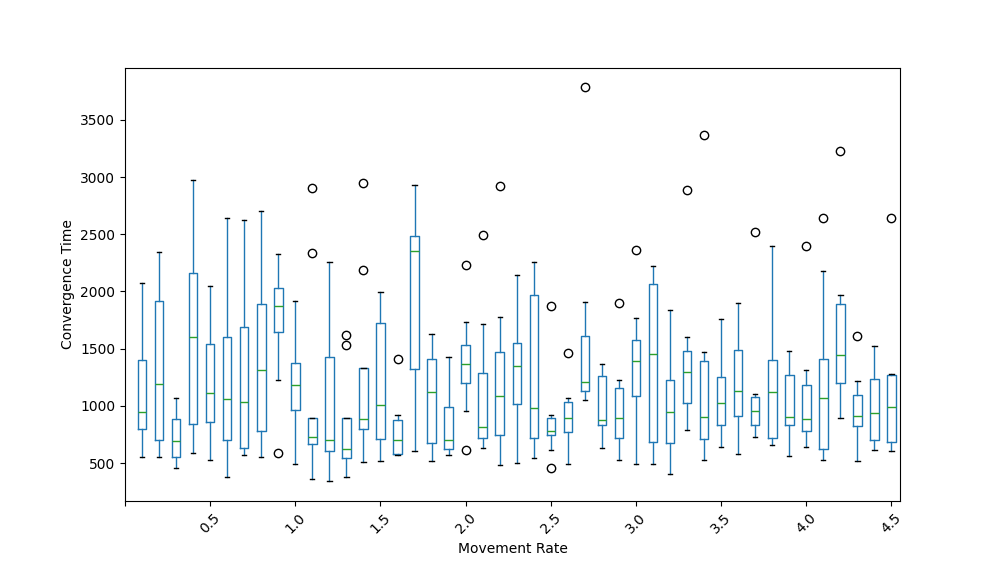}
	\caption{Box plot showing the distribution of convergence times for varying movement rates. Each box represents the range of convergence times observed for a specific movement rate, with outliers depicted as individual points. The movement rate here defines the maximum speed assigned to each node, which follows a normal distribution. This figure illustrates the complex relationship between node speed and convergence efficiency in the network.}
	\label{fig:movement-convergence-speed}
\end{figure}

The box plot in figure \ref{fig:movement-convergence-speed} highlights the variability and complexity of convergence times across different movement rates in the network. In these simulations, each node is assigned a random velocity, generated using a normal distribution with a defined velocity limit. Here, the movement rate refers to the velocity limit applied to the random velocities of nodes, providing insights into how node mobility influences convergence dynamics.
Notably, the convergence time exhibits significant variation at each speed limit, with wide interquartile ranges and numerous outliers. This variability suggests that network convergence is sensitive to node movement, as the continuous changes in node positions influence connectivity and, consequently, the time required to reach full convergence. However, a direct correlation between increased movement rate and faster or slower convergence time is not evident. 

The presence of multiple outliers across all movement rates in \ref{fig:movement-convergence-speed} further underscores the unpredictability of convergence, as certain configurations or node distributions lead to exceptionally high or low convergence times. These findings suggest that while node mobility impacts network adaptation, achieving optimal convergence likely requires considering additional parameters beyond movement rate alone. 
While increased mobility introduces challenges, the adaptive algorithm maintains robust convergence across a range of speeds.

\begin{figure*}[h!]
	\centering
	\begin{subfigure}[b]{0.48\textwidth}
		\centering
		\includegraphics[width=\textwidth]{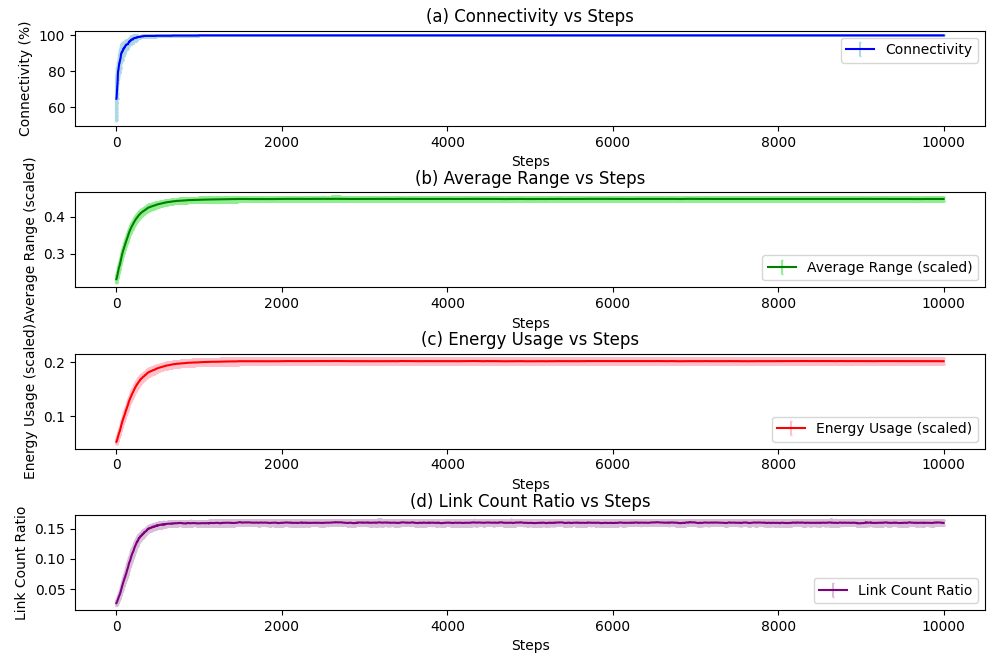}	
		\caption{Static 3D network model showing convergence to 100\% connectivity at a density of 0.05. The network adapts using local strategies to ensure robust connectivity, demonstrating similar performance as observed in the 2D model.}
		\label{fig:local-connectivity-3D}
	\end{subfigure}
	\hfill
	\begin{subfigure}[b]{0.48\textwidth}
		\centering
		\includegraphics[width=\textwidth]{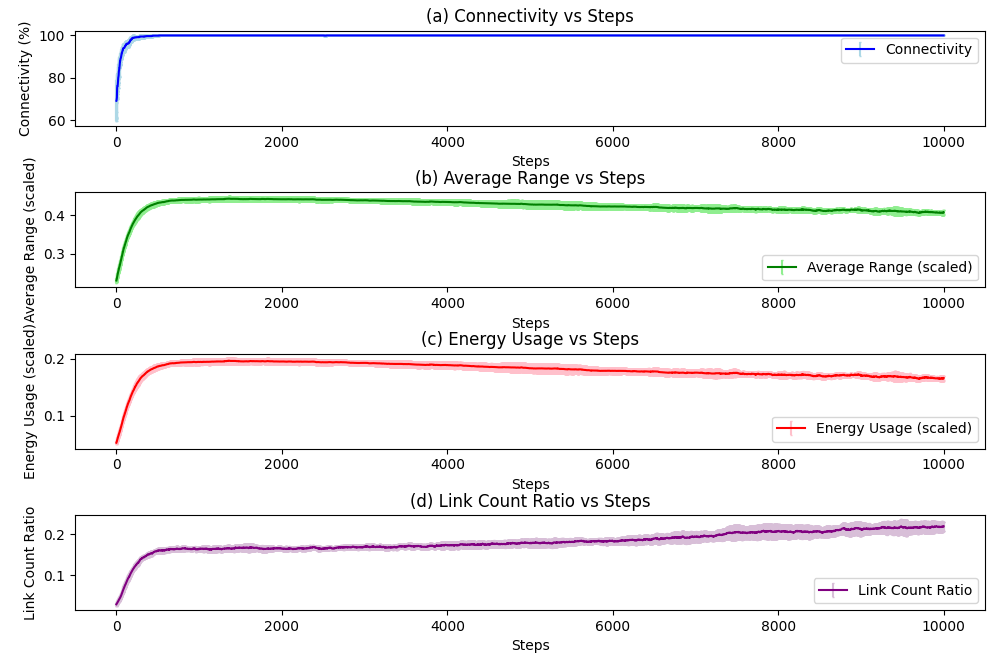}	
		\caption{Mobile 3D network model with node movements at a density of 0.05. The network achieves 100\% connectivity despite the dynamic topology, confirming the effectiveness of the proposed method in 3D environments.}
		\label{fig:movement-connectivity-3D}
	\end{subfigure}
	\caption{ Demonstration of the adaptability and robustness of the proposed method in a 3D network model. The figure consists of two sub-figures: (a) a static 3D network with local adaptation, and (b) a mobile 3D network with dynamic node movements. Both networks are initialized with the same conditions as the 2D model at a density of 0.05. The results confirm that the proposed method successfully achieves 100\% connectivity in 3D network scenarios, validating its generalizability across spatial dimensions.}
	\label{fig:3D}
\end{figure*}	

To demonstrate the generalization capability of our approach, we aim to validate its applicability to three-dimensional (3D) network models in both static and mobile scenarios. The results are depicted in figure \ref{fig:3D}. 
\ref{fig:local-connectivity-3D} illustrates that the method successfully achieves 100\% connectivity in a static 3D network with local adaptation, mirroring the performance of the 2D model. Similarly, \ref{fig:movement-connectivity-3D} demonstrates that the network maintains its adaptability in a mobile 3D scenario, overcoming challenges posed by dynamic node movements and topological changes. Stabilized transmission ranges and energy usage reinforce the algorithm's energy efficiency. 
The consistent performance across all main metrics highlights the generalizability of the approach without compromising connectivity or performance, even when transitioning from 2D to 3D scenarios.
These findings underscore the versatility of the approach and its potential applicability in diverse real-world environments requiring adaptive and resilient 3D network models.

%% file: sections/conclusion.tex
This study proposed a novel AI-driven self-organizing complex network model for achieving high connectivity, robustness, and energy efficiency in distributed networks. The integration of Hamiltonian-based methodologies with MLP decision-making enabled individual nodes to adapt their transmission power autonomously, optimizing global network performance while responding to local conditions. The results demonstrated the effectiveness of the proposed approach in both static and mobile scenarios, across 2D and 3D environments, confirming its scalability and applicability in diverse contexts.
Key findings highlight the network's ability to achieve 100\% connectivity with significant resilience to node failures, dynamic mobility, and varying network densities. The energy usage stabilized at optimal levels, reflecting the efficiency of the adaptive transmission power mechanism. This efficiency not only prolongs the lifetime of individual nodes but also enhances the overall network’s sustainability. Furthermore, the methodology's ability to balance connectivity and energy consumption underscores its potential for deployment in real-world applications, such as IoT systems, wireless sensor networks, and autonomous communication platforms.

Future work will focus on extending the model to incorporate more complex scenarios, including heterogeneous nodes with varying energy capacities and roles, as well as networks subjected to extreme environmental conditions.
Future work will explore incorporating the Hamiltonian concept into the deep learning architecture, aiming to leverage physics-guided machine learning approaches \cite{seyyedi2023machine}, having the potential to enhance the model's capabilities. Additionally, exploring the integration of advanced reinforcement learning techniques could further enhance the decision-making capabilities of individual nodes. Expanding the study to include real-world experiments and benchmarking against state-of-the-art approaches can also provide insights into the model's practical feasibility and performance. These advancements will pave the way for developing robust, scalable, and sustainable distributed networks capable of meeting the demands of modern communication systems.